\documentclass{ws-ijmpd}
\usepackage[super,compress]{cite}
\usepackage{hyperref}
\hypersetup{colorlinks=false}
\usepackage{myaasmacros}

\begin{document}
\newcommand{\be}{\begin{equation}}
\newcommand{\ee}{\end{equation}}
\newcommand{\bea}{\begin{eqnarray}}
\newcommand{\eea}{\end{eqnarray}}

\markboth{M.~S.~Pshirkov}
{May axion clusters be sources of fast radio bursts?}

%
\catchline{}{}{}{}{}

\title{May axion clusters be sources of fast radio bursts?}
\author{M.\,S.\,Pshirkov}
\address{ Sternberg Astronomical Institute, Lomonosov Moscow State University, Universitetsky prospekt 13, 119992, Moscow, Russia\\~\\
Institute for Nuclear Research of the Russian Academy of Sciences, 117312, Moscow, Russia\\~\\
Pushchino Radio Astronomy Observatory, 142290, Pushchino, Russia\\~\\
pshirkov@sai.msu.ru
}

\maketitle

\begin{history}
\received{Day Month Year}
\revised{Day Month Year}
\end{history}

\abstract{
Fast radio bursts can be caused by some phenomena related to ’new physics’.One of the most prominent candidates of the kind are axion Bose stars which can engender bursts when undergoing conversion into photons in magnetospheres of neutron stars or during their collapse. In this short research note an importance of three observational criteria is outlined, namely total energetic, $\mathcal{O}$(ms) duration and non-monochromatic shape of the spectrum.  It is shown that it is impossible to meet these criteria in simplified models of axion-neutron star interaction scenario and thorough investigation of more complex models which  involve back-reaction and non-linear effects are needed in order to explain FRBs within this scenario.
}
\ccode{14.80.Va, 97.60.Jd}

\vspace{1 cm}
Fast radio bursts (FRBs) is a new class of astrophysical phenomena, their origin is one of the major mysteries of the high-energy astrophysics. The first  FRB ('Lorimer burst', FRB010724) was found in  archival Parkes data  \cite{Lorimer2007}. Since then 17 more FRBs were discovered, thus firmly establishing new class of transient radio sources \cite{FRBcat}. The most striking features of these sources are  their large dispersion measures which solidly indicate their extragalactic origin and that implies that the energetics of these events is huge indeed -- it can be more than $10^{41}$~ergs and this amount is radiated during time span of less than  several ms.
Short duration suggests that the catastrophic processes are taking place in a rather limited volume and inevitably the vast majority  of scenarios include neutron stars (NSs) with their $\mathcal{O}(10^{6}~\mathrm{cm})$ size.
A  large number of models are striving to explain FRBs. Most popular astrophysical models are i) flares from NS with very strong magnetic fields-- magnetars \cite{Popov2010,Lyubarsky2014};  ii) supergiant pulses from energetic young pulsars, which are akin to well-known Crab giant pulses but exceed them in energetics by several orders of magnitude \cite{Cordes2016,Lyutikov2016,Popov2016};  iii)  FRBs from NS-NS mergers, when during the coalescence process a collapsar with magnetar-like field ($>10^{15}$~G) and very fast rotation ($\Omega\sim10^{4}$~Hz) can briefly form \cite{Pshirkov2010,Totani2013}. This list is by no means complete,   also there are many models invoking new physics as an explanation as well. One of the most popular kind of these models is using axions and, specifically, axion-NS interactions as sources of the FRBs.

The axion is a hypothetical particle which was introduced in order to explain strong CP problem \cite{PQ} and now it is one of the best motivated Dark Matter (DM) candidate. Axions are not completely  'dark' -- in strong magnetic fields they can be converted into photons of the same energy  via so-called Primakoff effect. E.g., if axions constituted considerable fraction of the DM, than this conversion could effectively take place in  magnetospheres of the neutron stars and resulting photons could be detected \cite{Pshirkov2009}.   Properties of axions can be used to explain FRBs as well. First of all, it was shown that certain fraction of axions may undergo Bose condensation  and, eventually, will reside in small dense Bose stars with characteristic mass $M_{bs}\sim10^{-12}~\mathrm{M_{\odot}}$ and radius $R_{bs}\sim300\left(\frac{10~\mathrm{\mu eV}}{m_a}\right)^2~\mathrm{km}$ \cite{Kolb1994,Chavanis2011a,Chavanis2011b,Tkachev2015}, where $m_a$ is the axion mass;   the number of such  Bose stars can be  very high in a Milky Way-sized galaxy.
If considerable amount of axions in a Bose star was converted into photons that could produce a bright short flare with  the total energy budget up to $E=M_{bs}c^2\sim10^{42}~$ergs which could  be detected   as a FRB from even cosmological distances \cite{Tkachev2015,Iwazaki1,Iwazaki2,Raby2016}.  This scenario is very attractive, it simultaneously points out right DM candidate (axion), gives its properties ($m_a\sim5~\mu\mathrm{eV}$) and explains mysterious FRBs, thus it merits careful investigation. 

There are several versions  of this  scenario. First, the conversion can take place in a vicinity of strongly magnetized neutron star, the  location depends on the exact  mechanism of the conversion. In \cite{Iwazaki1,Iwazaki2} it was suggested that the conversion proceeds at the very surface of the NS: coherent axionic field in presence of strong magnetic field gave rise to oscillating electric field which in turn forces electrons in dense  ($n_e\sim 10^{21}~\mathrm{cm^{-3}}$) 'atmosphere' radiate coherently.  Unfortunately, this mechanism can not operate effectively and thus can not explain FRBs. There are two main reasons for that: first, radio waves would not propagate in such dense plasma, though the presence of dense  plasma is absolutely necessary for the mechanism. The effect of free-free absorption is indeed very low in  presence of strong magnetic fields. However, it is relevant only for electromagnetic waves of very high frequency \cite{Potekhin2003}, $\nu>>\nu_{pl}$, where $\nu_{pl}=9\times10^{3}\sqrt{(n_e/\mathrm{1~cm^{-3}})}$~Hz is the plasma frequency which is close to $10^{14}$~Hz in our case. For ordinary radio waves with $\nu<<\nu_{pl}$  and electric field parallel to the external magnetic field the propagation is impossible exactly as in the case of unmagnetized plasma.   Second, the entire mechanism in \cite{Iwazaki2} violates the energy conservation law. Coherent emission (Eq. (10) of paper\cite{Iwazaki2}) can not produce larger luminosity  than all the individual dipoles combined ( Eq. (9), $L_{tot}=10^{33}\times10^{-9}~\mathrm{GeV~s^{-1}}\sim10^{21}~\mathrm{erg~s^{-1}}$, see \cite{Katz2014} for an extended discussion of coherent emission in FRBs). That only leaves conversion of axions to photons in strong external field of NS magnetospheres \cite{Pshirkov2009} as the  viable option \cite{Tkachev2015}.
Alternatively, FRBs can potentially be produced by intrinsic processes in Bose stars even in absence of any interaction with NSs: Axion Bose stars are unstable if their mass exceeds certain value, and collapse of such stars could lead to abundant photon emission due to stimulated axion decay $a\rightarrow\gamma\gamma$ \cite{Tkachev1986,Tkachev2015,Levkov2016}.

There are several main features of the FRBs which any successful scenario must reproduce:
\begin{itemize}
\item(i) their huge energetics, if we are dealing with true extragalactic phenomena;
\item(ii) their exceedingly short intrinsic duration, $\mathcal{O}$(ms); 
\item(iii) their spectral shape: most of the bursts were detected at frequencies around 1.4 GHz in a several hundred MHz bandwidth observations, however, in the end of 2015 it was announced that  FRB110523 was detected by the Green Bank radiotelescope in 0.7-0.9 GHz band \cite{Masui2015}. Of course, like, e.g. gamma-ray bursts, FRBs can comprise heterogeneous types of sources and with the limited knowledge of their spectral properties at the moment, the minimal requirement for any scenario is $\Delta f/f>10\%$, where $\Delta f $ is the spectral width of the  signal with the central frequency $f$;
\end{itemize}

As shown in \cite{Iwazaki1,Tkachev2015}, the first condition can be  met in axion Bose star scenarios, providing that the substantial part of its mass will be converted into photons. The conversion efficiency in  magnetospheres of NS with large magnetic fields --  magnetars -- can be of order unity\cite{Pshirkov2009}. In case of Bose star collapse it is expected that  a large  fraction of its mass will go into photons.

It is much more difficult to satisfy  the second condition in the NS-related model. The reason is very simple --   axion cluster is an extremely fragile object that is barely bound. In presence of  strong tidal field  of the neutron star  it will be  destroyed at the radius of tidal destruction $r_{td}$:
\begin{equation}
	r_{td}=R_{bs}\left(\frac{M_{bs}}{M_{NS}}\right)^{-1/3}=3\times10^{11}~\mathrm{cm}.
	\label{tidal_radius}
	\end{equation}
	
At this radius the cluster will be falling at $v_{td}=\sqrt{v_{\infty}^2+\frac{2GM_{NS}}{r_{td}}}=4.6\times10^{7}~\mathrm{cm~s^{-1}}$, where $v_{\infty}=3\times 10^{7}~\mathrm{cm~s^{-1}}$ is the asymptotical velocity of the cluster at infinity and $M_{NS}=1.4~M_{\odot}$.
The total duration of the transient event will be defined by a simple  time of crossing $t_{td}=2R_{bs}/v_{td}=1.3$~s which is much larger than the typical  duration of FRBs $\mathcal{O}$(ms). The latter short duration could only be reached if the cluster arrived at the NS surface without any disruption, retaining its initial radius, e.g. if axions could actually form denser configurations due to balance between mean-field pressure of axion condensate and gravitational forces  \cite{Braaten2016} and  density could reach nuclear levels. In this case these objects would be completely immune to even NS tidal fields and would arrive to the NS surface intact, perfectly capable to engender a FRB. 
Also there is a possibility that this long timescales can be just an artefact of oversimplified approach: in reality the conversion of this huge amount of energy will definitely be non-linear process with strong backreaction, i.e. the properties of  magnetosphere can be quicky altered in the first moments of the interaction and that in turn will quench any further conversion, thus effectively enforcing  much shorter timescales. This question will be more thoroughly studied elsewhere.

It is not so easy  for axion-related scenarios  to reproduce  spectral properties of the FRBs  either: the simplified scenarios predict almost monochromatic signal which frequency is defined by the axion mass. It can be slightly widened because of Doppler and relativistic corrections but the relative  width can not be much larger than 10~\%.  On the other hand, much more complex behaviour, especially in the case of Bose star collapse scenario, can be  expected e.g., non-linear effects lead to imminent widening of emitted spectrum and even result in a  power law spectrum\cite{Khlebnikov1996,Micha2004}.

Even if an accurate account for backreaction and non-linearity of  Bose star interactions will not be able to produce needed short timescales and wide spectral features, still, axions can be main constituent of the DM and then they   can  form axion clusters. Thus, search for distinctive astrophysical phenomena that can be possibly related with interactions of this clusters with NS magnetic fields is very important for the whole field of  astroparticle physics. If axion Bose stars are related to FRB, there can be a new class of transient  with a time duration around several seconds and a  narrow spectrum that will be its main distinctive feature.
Future observations with wider spectral coverage and larger field of view will allow to test all these scenarios and put strong constraints on axion content of the DM.

\newpage
\paragraph*{Acknowledgements.} 
The author wants to thank S. Troitsky, A. Panin and  G. Rubtsov for valuable comments. The author would also like to thank the anonymous referee whose comments and suggestions helped to significantly improve the quality of the paper. The work of the author is supported by  the Russian Science Foundation grant 14-12-01340. 

\begin{thebibliography}{10}

\bibitem{Lorimer2007}
D.~R. Lorimer, M.~Bailes, M.~A. McLaughlin, D.~J. Narkevic and F.~Crawford,
  {\em Science} {\bf 318}  (2007)   777,
  \href{http://arxiv.org/abs/0709.4301}{{\ttfamily arXiv:0709.4301
  [astro-ph]}}.

\bibitem{FRBcat}
E.~Petroff, E.~D. Barr, A.~Jameson, E.~F. Keane, M.~Bailes, M.~Kramer,
  V.~Morello, D.~Tabbara and W.~van Straten  (2016)
  \href{http://arxiv.org/abs/1601.03547}{{\ttfamily arXiv:1601.03547
  [astro-ph.HE]}}.

\bibitem{Popov2010}
S.~B. {Popov} and K.~A. {Postnov}, { {Hyperflares of SGRs as an engine for
  millisecond extragalactic radio bursts}}, in {\em Evolution of Cosmic Objects
  through their Physical Activity\/},  eds. H.~A. {Harutyunian}, A.~M.
  {Mickaelian} and Y.~{Terzian} (November 2010), pp. 129--132.
\newblock \href{http://arxiv.org/abs/0710.2006}{{\ttfamily arXiv:0710.2006}}.

\bibitem{Lyubarsky2014}
Y.~{Lyubarsky}, {\em \mnras} {\bf 442} (July 2014) L9,
  \href{http://arxiv.org/abs/1401.6674}{{\ttfamily arXiv:1401.6674
  [astro-ph.HE]}}.

\bibitem{Cordes2016}
J.~M. {Cordes} and I.~{Wasserman}, {\em \mnras} {\bf 457} (March 2016) 232,
  \href{http://arxiv.org/abs/1501.00753}{{\ttfamily arXiv:1501.00753
  [astro-ph.HE]}}.

\bibitem{Lyutikov2016}
M.~Lyutikov, L.~Burzawa and S.~B. Popov  (2016)
  \href{http://arxiv.org/abs/1603.02891}{{\ttfamily arXiv:1603.02891
  [astro-ph.HE]}}.

\bibitem{Popov2016}
S.~B. {Popov} and M.~S. {Pshirkov}, {\em \mnras} {\bf 462} (October 2016) L16,
  \href{http://arxiv.org/abs/1605.01992}{{\ttfamily arXiv:1605.01992
  [astro-ph.HE]}}.

\bibitem{Pshirkov2010}
M.~S. {Pshirkov} and K.~A. {Postnov}, {\em \apss} {\bf 330} (November 2010) 13,
  \href{http://arxiv.org/abs/1004.5115}{{\ttfamily arXiv:1004.5115
  [astro-ph.HE]}}.

\bibitem{Totani2013}
T.~{Totani}, {\em \pasj} {\bf 65} (October 2013)  ~12,
  \href{http://arxiv.org/abs/1307.4985}{{\ttfamily arXiv:1307.4985
  [astro-ph.HE]}}.

\bibitem{PQ}
R.~D. {Peccei} and H.~R. {Quinn}, {\em Physical Review Letters} {\bf 38} (June
  1977) 1440.

\bibitem{Pshirkov2009}
M.~S. {Pshirkov} and S.~B. {Popov}, {\em Soviet Journal of Experimental and
  Theoretical Physics} {\bf 108} (March 2009) 384,
  \href{http://arxiv.org/abs/0711.1264}{{\ttfamily arXiv:0711.1264}}.

\bibitem{Kolb1994}
E.~W. {Kolb} and I.~I. {Tkachev}, {\em \prd} {\bf 50} (July 1994) 769,
  \href{http://arxiv.org/abs/astro-ph/9403011}{{\ttfamily astro-ph/9403011}}.

\bibitem{Chavanis2011a}
P.-H. {Chavanis}, {\em \prd} {\bf 84} (August 2011)   043531,
  \href{http://arxiv.org/abs/1103.2050}{{\ttfamily arXiv:1103.2050}}.

\bibitem{Chavanis2011b}
P.-H. {Chavanis} and L.~{Delfini}, {\em \prd} {\bf 84} (August 2011)   043532,
  \href{http://arxiv.org/abs/1103.2054}{{\ttfamily arXiv:1103.2054
  [astro-ph.CO]}}.

\bibitem{Tkachev2015}
I.~I. {Tkachev}, {\em Soviet Journal of Experimental and Theoretical Physics
  Letters} {\bf 101} (January 2015) 1,
  \href{http://arxiv.org/abs/1411.3900}{{\ttfamily arXiv:1411.3900
  [astro-ph.HE]}}.

\bibitem{Iwazaki1}
A.~{Iwazaki}, {\em \prd} {\bf 91} (January 2015)   023008,
  \href{http://arxiv.org/abs/1410.4323}{{\ttfamily arXiv:1410.4323 [hep-ph]}}.

\bibitem{Iwazaki2}
A.~{Iwazaki}, {\em ArXiv e-prints}  (December 2015)
  \href{http://arxiv.org/abs/1512.06245}{{\ttfamily arXiv:1512.06245
  [hep-ph]}}.

\bibitem{Raby2016}
S.~{Raby}, {\em ArXiv e-prints}  (September 2016)
  \href{http://arxiv.org/abs/1609.01694}{{\ttfamily arXiv:1609.01694
  [hep-ph]}}.

\bibitem{Potekhin2003}
A.~Y. {Potekhin} and G.~{Chabrier}, {\em \apj} {\bf 585} (March 2003) 955,
  \href{http://arxiv.org/abs/astro-ph/0212062}{{\ttfamily astro-ph/0212062}}.

\bibitem{Katz2014}
J.~I. {Katz}, {\em \prd} {\bf 89} (May 2014)   103009,
  \href{http://arxiv.org/abs/1309.3538}{{\ttfamily arXiv:1309.3538
  [astro-ph.HE]}}.

\bibitem{Tkachev1986}
I.~I. {Tkachev}, {\em Soviet Astronomy Letters} {\bf 12} (October 1986) 305.

\bibitem{Levkov2016}
D.~G. {Levkov}, A.~G. {Panin} and I.~I. {Tkachev}, {\em ArXiv e-prints}
  (September 2016) \href{http://arxiv.org/abs/1609.03611}{{\ttfamily
  arXiv:1609.03611}}.

\bibitem{Masui2015}
K.~Masui {\em et~al.}, {\em Nature} {\bf 528}  (2015)   523,
  \href{http://arxiv.org/abs/1512.00529}{{\ttfamily arXiv:1512.00529
  [astro-ph.HE]}}.

\bibitem{Braaten2016}
E.~{Braaten}, A.~{Mohapatra} and H.~{Zhang}, {\em Physical Review Letters} {\bf
  117} (September 2016)   121801,
  \href{http://arxiv.org/abs/1512.00108}{{\ttfamily arXiv:1512.00108
  [hep-ph]}}.

\bibitem{Khlebnikov1996}
S.~Y. {Khlebnikov} and I.~I. {Tkachev}, {\em Physical Review Letters} {\bf 77}
  (July 1996) 219, \href{http://arxiv.org/abs/hep-ph/9603378}{{\ttfamily
  hep-ph/9603378}}.

\bibitem{Micha2004}
R.~{Micha} and I.~I. {Tkachev}, {\em \prd} {\bf 70} (August 2004)   043538,
  \href{http://arxiv.org/abs/hep-ph/0403101}{{\ttfamily hep-ph/0403101}}.

\end{thebibliography}


\end{document}